\documentclass[preprintnumbers,twocolumn,aps,amsmath,amssymb,showpacs,nofootinbib,superscriptaddress]{revtex4-2}

\usepackage{amsmath}
\usepackage{graphicx,enumerate,xcolor,enumitem,xspace,braket,mathrsfs} 
\usepackage[colorlinks,citecolor=blue]{hyperref}
\usepackage{comment}
\usepackage{float}
\usepackage{booktabs}
\usepackage{cleveref} 
\usepackage{braket}
\usepackage[normalem]{ulem}

\begin{document}
\preprint{PITT-PACC-2608}

\title{Leggett-Garg Inequality Violation in Muon $g-2$ Experiments}

\author{Brian Batell}
\email{batell@pitt.edu}
\affiliation{PITT PACC, Department of Physics and Astronomy,\\ University of Pittsburgh, 3941 O’Hara St., Pittsburgh, PA 15260, USA}

\author{Morgan Cassidy}
\email{MEC400@pitt.edu}
\affiliation{PITT PACC, Department of Physics and Astronomy,\\ University of Pittsburgh, 3941 O’Hara St., Pittsburgh, PA 15260, USA}

\author{Kun Cheng}
\email{kun.cheng@pitt.edu}
\affiliation{PITT PACC, Department of Physics and Astronomy,\\ University of Pittsburgh, 3941 O’Hara St., Pittsburgh, PA 15260, USA}

\begin{abstract}
   We present the first study of Leggett-Garg inequality violation in polarized muon spin precession. 
   We formulate a procedure to reconstruct temporal correlators of the longitudinal muon polarization from measured time-dependent muon decay spectra and apply it to publicly available Fermilab Muon $g-2$ data corresponding to approximately $10$ billion muon decays.
  Using a simplified model of the detector acceptance and efficiency, the Leggett-Garg inequality is found to be violated with a single-bin significance of $5.5\sigma$, while combining neighboring bins further increases the significance. While our analysis is limited by systematic uncertainties associated with the detector modeling, a dedicated experimental analysis could reduce these uncertainties toward the statistical level, $\mathcal{O}(10^{-3})$, potentially enabling one of the most precise measurements of temporal quantum correlations.
\end{abstract}

\maketitle

\section{Introduction}

The principle of superposition and the non-commutative properties of quantum mechanics lead to counterintuitive quantum correlations, such as entanglement~\cite{HORODECKI1997333,PhysRevLett.80.2245} and Bell inequality violation~\cite{PhysicsPhysiqueFizika.1.195,Clauser:1969ny}, that do not exist in classical systems. 
Among the many ways to characterize the differences between quantum and classical correlations, Bell inequalities occupy a central role by providing tests of local realism based on correlations between spatially separated systems.

Quantum correlations can arise not only in spatially separated systems, but also in a single evolving system. 
The Leggett-Garg inequality (LGI), also known as the temporal Bell inequality, focuses on the quantum correlations between the past and future of an evolving system~\cite{Leggett:1985zz}.  
Instead of assuming locality as in the Bell inequality, the LGI assumes that non-invasive measurements can be performed on an evolving system at different times, leading to an analogous inequality that must be respected by realist descriptions but can be violated in quantum systems. 

More recently, there has been growing interest in exploring aspects of quantum information in high energy experiments, mostly focusing on quantum correlations of particle pairs~\cite{ATLAS:2023fsd,CMSCollaboration_2024,Afik:2020onf,Fabbrichesi:2021npl,Barr:2021zcp,Severi:2021cnj,Afik:2022kwm,Afik:2022dgh,Han:2023fci,Aguilar_Saavedra_2022,Dong_2024,Cheng:2023qmz,Cheng:2024btk,Han:2024ugl,Cheng:2024rxi,Altakach:2022ywa,Ehataht:2023zzt,Ma:2023yvd,Fabbrichesi:2024wcd,Han:2025ewp,Zhang:2025mmm,Ai:2025wnt,Barr:2022wyq,Ashby-Pickering:2022umy,Aguilar-Saavedra:2022wam,Fabbrichesi:2023cev,Fabbri:2023ncz,Bi:2023uop,Morales:2023gow,Bernal:2024xhm,Grossi:2024jae,Goncalves:2025qem,Goncalves:2025xer,Afik:2025grr,Cheng:2025cuv,Lin:2025eci,Guo:2026yhz,Fang:2026ddi,Qi:2025onf,Cheng:2025zaw,Zhang:2026nwm}. 
These high energy environments also provide attractive settings to study the LGI. 
Systems such as mesons~\cite{Naikoo:2018amb,Naikoo:2018vug,Blasone:2024jur,Cheng:2025zcf} and neutrinos~\cite{Gangopadhyay:2013aha,Alok:2014gya,Banerjee:2015mha,Formaggio:2016cuh,Fu:2017hky,Gangopadhyay:2017nsn,Naikoo:2019eec,Wang:2022tnr,Alam:2026bxn} provide especially useful examples, since their time evolution exhibits coherent quantum oscillations.
Among these systems, the LGI has been widely investigated in neutrino oscillations. 
Given an identically prepared neutrino ensemble in the same flavor state, the energy difference between two neutrino events can be translated into oscillation time difference, allowing a reconstruction of the LGI under the stationarity assumption.

In this work, we study the LGI in the spin precession of high energy polarized muons in muon $g-2$ experiments, with particular relevance to the programs at Fermilab and J-PARC~\cite{Muong-2:2015xgu,Iinuma:2011zz}.  
This system is especially well suited for a LGI study: the initial muon polarization is well characterized, the spin evolution takes place in a highly controlled storage ring environment, and the time-dependent decay spectrum provides access to the spin-precession signal.
Since the muon $g-2$ measurement is among the most precise quantum measurements in high energy physics, providing a determination of the muon spin precession frequency with a relative uncertainty of $10^{-7}$~\cite{Muong-2:2025xyk}, it offers a promising route toward a precision study of temporal quantum correlations.

We develop a method to reconstruct the LGI from the time-dependent muon decay spectrum reported in muon $g-2$ experiments.
Utilizing two oscillation periods of the muon decay spectrum data from the Fermilab Muon $g-2$ experiment~\cite{Muong-2:2015xgu}, corresponding to around 10 billion muon events~\cite{LaBounty:2024epq}, we determine the muon spin at different times and use the resulting temporal correlations to construct the third-order Leggett-Garg variable $K_3$.
As a first illustration, we use a simplified model to describe detector effects such as resolution and acceptance. 
We treat the residual difference between the experimental data and our prediction of the decay spectrum as a systematic uncertainty associated with this simplified detector modeling. Nevertheless, even with this conservative treatment, we find that the LGI is violated with a single-bin significance of  $5.5\sigma$. Our study thus provides the first determination of LGI violation with oscillating muons. The statistical uncertainty is only $\mathcal{O}(10^{-3})$, and a dedicated reanalysis of the existing data with full experimental information by the Fermilab Muon $g-2$ collaboration could reduce the total uncertainty toward the statistical level, potentially enabling one of the most precise probes of temporal quantum correlations.

\section{Leggett-Garg inequality}

Consider a dichotomic observable $\hat Q$ (with outcomes $\pm 1$) that can be measured at different times in an evolving system. The autocorrelation function between measurements at times $t_i$ and $t_j$ is
\begin{equation}\label{eq:CijQiQj}
    C_{ij} = \left\langle \hat Q(t_i) \hat Q(t_j) \right\rangle.
\end{equation}
If realism holds and measurements are noninvasive, then each $\hat Q(t_i)$ in Eq.~\eqref{eq:CijQiQj} has a definite value, and a measurement at one time does not influence a measurement at a later time.
Therefore, similar to the construction of Bell inequalities, the following Leggett-Garg inequality is satisfied~\cite{Leggett:1985zz},
\begin{align}\label{eq:K3}
    K_3&\equiv C(t_1,t_2)+C(t_2,t_3)-C(t_1,t_3) \leq 1 \, .
\end{align}
In quantum mechanics, the maximum allowed value of $K_3$ is instead $3/2$, and the violation of the LGI reflects the ``quantumness" of the system's evolution.

The original derivation of the LGI assumed that measurements of $\hat Q$ at various times are made in a non-invasive manner. 
A modified version of the LGI can instead be derived within the ``stationarity'' framework~\cite{PhysRevA.52.R2497,Waldherr:2011kgt,Zhou:2015bbe,Emary:2013wfl}.
Stationarity assumes that the autocorrelation $C(t_i,t_j)$ depends only on the time difference $\Delta t_{ij} \equiv t_j - t_i$ between measurements, i.e.,  $C(t_i,t_j)=C(\Delta t_{ij})$. Considering the case of equally spaced measurement times,  $\Delta t_{12} = \Delta t_{23} = \Delta t$, the three-time LGI in Eq.~\eqref{eq:K3Deltat} takes the simple form~\cite{PhysRevA.52.R2497,Zhou:2015bbe}~\footnote{In the stationarity-based formulation, the LGI applies to macrorealist descriptions under the additional assumptions of time-translation-invariant transition probabilities, Markovian dynamics, and preparation of the system in a specified initial state~\cite{Emary:2013wfl}.}
\begin{equation}\label{eq:K3Deltat}
    K_3 = 2C(\Delta t) - C(2\Delta t) \leq 1 \, .
\end{equation}
The autocorrelation function may then be written in terms of conditional probabilities as~\cite{PhysRevA.52.R2497}
\begin{equation}\label{eq:CfromConditionalP}
    C(\Delta t)= P_{++}^{\Delta t} -  P_{+-}^{\Delta t}  =  P_{--}^{\Delta t} -  P_{-+}^{\Delta t} \, ,
\end{equation}
where $P_{\alpha\beta}^{\Delta t}$ is the probability that the measurement outcome at time $t_0+\Delta t$ is $\beta$, conditioned on the outcome at time $t_0$ being $\alpha$.
The LGI can then be reconstructed accordingly. 

In this work, we use the LGI as a diagnostic of temporal quantum correlations in the muon spin precession system. 
Temporal correlators of the longitudinal muon polarization are reconstructed from a highly polarized muon ensemble prepared in a specified initial state.
The longitudinal muon polarization is inferred from the measured decay positron energy spectra using the standard spin-dependent distribution for weak muon decay, analogous to spin correlation measurements with unstable particles at colliders~\cite{Afik:2020onf,Fabbrichesi:2021npl,Barr:2021zcp}. 
Therefore, just as collider studies of entangled particle pairs do not provide loophole-free tests of local realism when the relevant spin observables are inferred from decay distributions~\cite{Abel:1992kz,Li:2024luk,Bechtle:2025ugc,Abel:2025skj,Low:2025aqq,Aguilar-Saavedra:2026rsx}, our analysis does not constitute a general loophole-free test of macrorealism.\,\footnote{We note that it may be possible to exclude certain restricted classes of macrorealist descriptions under additional assumptions. For related discussions in the context of collider tests of local realism, see Refs.~\cite{Abel:1992kz,Li:2024luk,Bechtle:2025ugc,Aguilar-Saavedra:2026rsx}.}
The major focus of our analysis is to provide a precision probe of temporal quantum correlations in polarized muon precession.

\section{Muon $g-2$ experiments}

In the muon $g-2$ experiment, positively charged muons are produced from $\pi^+\to \mu^+\nu_\mu$ decays. Since the pion is a scalar and the neutrino is left-handed, the $\mu^+$ is fully polarized in the pion rest frame, with its spin anti-aligned with its momentum.
In the Fermilab beamline, the pions decay in flight, and the momentum selection preferentially accepts the highest momentum decay muons, which are emitted nearly along the pion direction and therefore retain a large net polarization of approximately $95\%$.
These polarized muons are then transported and injected into the storage ring with a precisely known magnetic field. At the magic momentum $3.09$ GeV, the electric field contribution to the relative spin precession is canceled, so the muon spin precesses relative to its momentum at the anomalous precession frequency set by the anomalous magnetic moment. This spin precession is measured through the parity-violating decay $\mu^+\to e^+\nu_e\bar\nu_\mu$, whose positron energy and angular distributions are correlated with the muon spin. In the muon rest frame, higher energy positrons are preferentially emitted along the muon spin direction. Consequently, in the laboratory frame, the number of detected high energy positrons oscillates as the muon spin precesses, allowing the anomalous precession frequency to be extracted.

\begin{figure}
    \centering
    \includegraphics[scale=0.53]
    {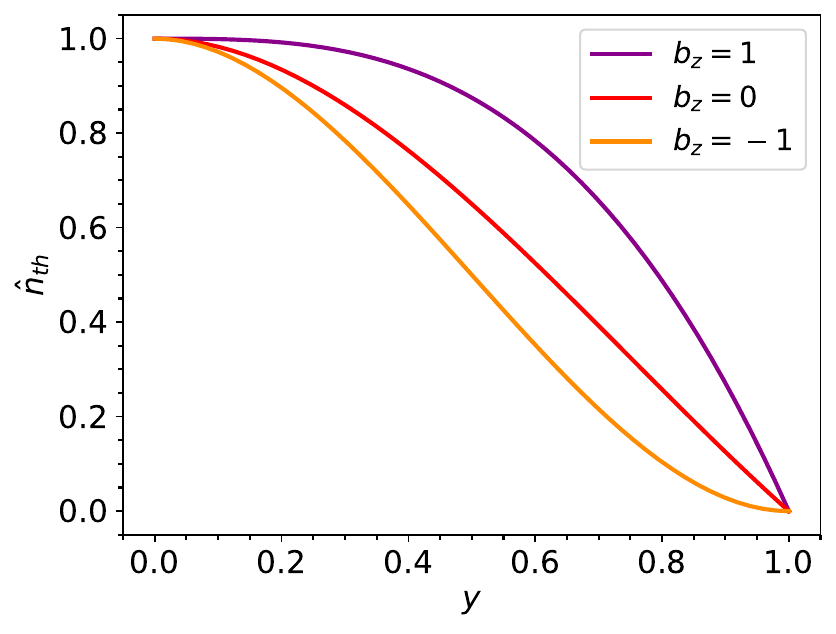}
    \caption{
    Normalized energy spectrum of the decay positrons as a function of $y = E/E_{\rm max}$ for longitudinal polarizations $b_{z} = + 1, 0, -1$ (respectively purple, red, orange).  
    } 
    \label{fig:Edistri_TH}
\end{figure}

More explicitly, consider a boosted muon moving along the $z$ direction with energy $E_\mu$ and longitudinal polarization $b_z=\langle\sigma_z\rangle$. The energy spectrum of the decay positron is given by~\cite{Muong-2:2015xgu}
\begin{align}\label{eq:positronDistribution}
    n_{\rm th}(E,b_z) =   N_{\rm th}(E)\Big[1+b_zA_{\rm th}(E)\Big] \, ,
\end{align}
where $E$ is the positron energy and 
\begin{subequations}\label{eq:NeAe_TH}
\begin{align}
    N_{\rm th}(E)&\propto (y-1)(4y^2-5y-5) \, ,  \\
    A_{\rm th}(E)&= \frac{-8y^2+y+1}{4y^2-5y-5} \, ,
\end{align}
\end{subequations}
with $y=E/E_{\rm max}$,  and the maximum energy of the positron is $E_{\rm max} \approx E_\mu$. Here, $N_{\rm th}(E)$ is the overall positron spectrum averaging over muon spin, and $A_{\rm th}(E)$ is the decay asymmetry.
The $b_z$ dependence of the positron energy spectrum is shown in Fig.~\ref{fig:Edistri_TH}, where, following the presentation of the experimental results, we normalize all distributions to their value in the first energy bin, denoted as $\hat n$.
Taking the quantum mechanical prediction of spin precession into account, the time dependence of the energy spectrum of decay positrons is given by 
\begin{align}\label{eq:positronDistribution_tdependent}
    n_{\rm th}(E,b_z(t))= N_{\rm th}(E) \Big[1+ b_{\rm max}\cos (\omega_a t+\phi) A_{\rm th}(E)\Big], \nonumber
\end{align}
where we have omitted the exponential decay factor 
and we take $b_{\rm max}=1$  for simplicity in the following. 

For the original purpose of measuring the anomalous magnetic moment in the muon $g-2$ experiment, the key observable is the anomalous precession frequency $\omega_a$. This can be obtained by constructing a time series of detected positrons, typically with an energy threshold or energy-dependent asymmetry weighting, and fitting the resulting oscillation in the positron rate. It is not necessary to reconstruct the muon spin polarization as a function of time by fitting the full positron energy spectrum in each time bin.
In contrast, reconstructing the LGI via Eq.~\eqref{eq:K3Deltat} requires the conditional probabilities for the muon to be spin-up or spin-down along its momentum direction at different time separations, given a specified spin state at time $t_0$.
To obtain these probabilities, we fit the full positron energy spectrum in Eq.~\eqref{eq:positronDistribution} at different decay times and extract the time-dependent longitudinal polarization $b_z(t)$. 
The corresponding probabilities are
\begin{equation}
\label{eq:Ppm}
    P_{\pm}(t) = \frac{1\pm b_z(t)}{2} \, .
\end{equation}
Choosing a reference time $t_0$ such that the muon ensemble is in the spin-down state, the conditional probabilities $P_{-+}^{t-t_0}$ and $P_{--}^{t-t_0}$
in Eq.~\eqref{eq:CfromConditionalP} can then be reconstructed from the extracted polarization $b_z(t)$.

\section{Reconstructing the LGI}

\begin{figure}
    \centering
    \includegraphics[scale=0.5]{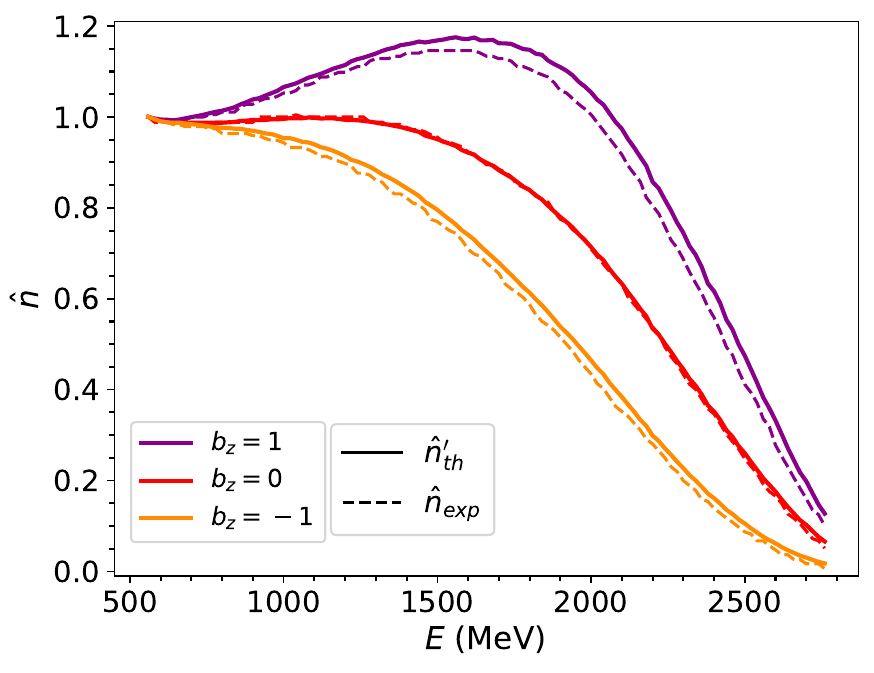}
    \caption{Comparison of measured decay positron normalized energy distributions with the theory prediction including detector response (\ref{eq:efficiency},\ref{eq:TH_n_eff})
    for $b_z =+ 1, 0, -1$. 
    Dashed lines correspond to measured values at 44.5 $\mu$s, 45.5 $\mu$s, and 46.6 $\mu$s. 
    }
    \label{fig:Edistri_EXP}
\end{figure}

In practice, detector effects can distort the measured positron energy spectrum relative to the theoretical prediction. 
As shown in Refs.~\cite{Muong-2:2015xgu,LaBounty:2024epq}, both the experimentally determined spin-averaged spectrum $N_{\rm exp}(E)$ and the energy-dependent asymmetry $A_{\rm exp}(E)$ differ from  Eq.~\eqref{eq:NeAe_TH}. 
The ratio $N_{\rm exp}(E)/N_{\rm th}(E)$ gives an effective energy-dependent detector efficiency for the spin-averaged spectrum, 
\begin{equation}\label{eq:efficiency0}
    \epsilon_0( E)  \propto \frac{N_{\exp}(E)}{N_{\rm th}(E)} \, .
\end{equation}
The difference between $A_{\rm exp}(E)$ and $A_{\rm th}(E)$ further indicates that the detector response also depends on $b_z$.

Therefore, as a simplified model, we assume that the detector efficiency is a linear function of polarization,
\begin{equation}\label{eq:efficiency}
    \epsilon(E, b_z)=\epsilon_0(E)+ b_z \epsilon_1(E) \, .
\end{equation}
The predicted spectrum is then given by
\begin{equation}\label{eq:TH_n_eff}
    n_{\rm th}'(E,b_z) = n_{\rm th}(E,b_z) \times \epsilon(E,b_z) \, .
\end{equation}
The $b_z$ dependent correction, $\epsilon_1(E)$, is determined from the reported values of $A_{\rm exp}(E)$ in Fig.~5.6 of Ref.~\cite{LaBounty:2024epq} by solving  
\begin{equation}
    \frac{ n_{\rm th}'(E,1)- n_{\rm th}'(E,-1) }{ n_{\rm th}'(E,1)+ n_{\rm th}'(E,-1) } = A_{\rm exp}(E),\nonumber
\end{equation}
from which we obtain
\begin{equation}\label{eq:efficiency1}
    \epsilon_1(E) 
    =\epsilon_0(E)\frac{ A_{\rm th}(E)-A_{\rm exp}(E)}{A_{\rm th}(E)A_{\rm exp}(E)-1} \, .
\end{equation}

In Fig.~\ref{fig:Edistri_EXP}, we show the energy distributions of positrons from muon decay for different muon polarizations after including detector effects.
The solid lines show the theoretical prediction from our simplified model in Eq.~\eqref{eq:TH_n_eff},
while the dashed lines show the experimentally measured positron spectra.
Although Eq.~\eqref{eq:efficiency} is only a simple linear parametrization, the theoretical spectra including detector effects (\ref{eq:TH_n_eff}) reproduce the overall shapes and magnitudes of the experimental spectra well, as shown in Fig.~\ref{fig:Edistri_EXP}. Small residual differences remain visible between the model prediction and the data. Because the statistical uncertainties in the data are much smaller than these differences, the residuals are treated below as systematic uncertainties associated with our simplified detector response model.

We digitize the positron energy spectra provided in Fig.~5.7 of Ref.~\cite{LaBounty:2024epq} at different decay times over two oscillation periods, beginning at $44.5~\mu{\rm s}$. 
The muon decay events are divided into 29-30 bins of decay time per oscillation period. In each time bin the positron energy spectrum is given by the counts in 111 energy bins spanning 560 to 2780 MeV.
We use $n_{\rm th}'(E,b_z) $ in Eq.~\eqref{eq:TH_n_eff} as a template to fit the extracted spectra and determine the muon polarization $b_z$ at each decay time. 
As in Fig.~\ref{fig:Edistri_EXP}, we normalize both our template $n_{\rm th}'(E,b_z)$ and the experimental data to the value of the first energy bin for the fitting, denoted as $\hat{n}$. 
At each decay time $t$, we construct the $\chi^2$ test statistic,
\begin{equation}\label{eq:chi2}
    \chi^2 = \sum_i \left( \frac{\hat n_{\exp}(E_i, t) - \hat n_{\rm th}'(E_i ,b_z)}{\Delta \hat n_i} \ \right)^2,
\end{equation}
where $\hat{n}_{\rm exp}(E_i,t)$ is the measured number of events in the $i$th energy bin, and $\Delta \hat n_i$ is the corresponding statistical uncertainty.

Since Eq.~\eqref{eq:efficiency} is a simplified model that does not fully characterize the detector response, we treat the residual discrepancies in Fig.~\ref{fig:Edistri_EXP} as systematic uncertainties on our theory template $\hat n'_{\rm th}(E_i,b_z)$.
Specifically, we define $\Delta(E_i,t)=\hat{n}_{\rm exp}(E_i,t)-\hat{n}_{\rm th}'(E_i,b_{\rm max}\cos(\omega_a t + \phi))$. We emphasize that the quantum mechanical form of precession is used only to estimate the systematic uncertainty, but is not assumed in the fit for $b_z(t)$.

Systematic uncertainties associated with finite time resolution are accounted for by varying the initial time of the theory template by the width of one time bin and choosing the maximum value of $\Delta$. 
To incorporate systematic uncertainties in our fitting, we follow the offset method in Ref.~\cite{Botje_2002} and modify the $\chi^2$ in Eq. (\ref{eq:chi2}) to
\begin{align}
 \chi^2 = &\sum_i \left( \frac{\hat n_{\exp}(E_i, t) - \hat n_{\rm th}'(E_i ,b_z) - s\Delta(E_i, t)}{\Delta \hat n_i}  \right)^2 \, . \nonumber
\end{align}
Here, the parameter $s$ takes on the values $\pm1$ and only statistical errors are taken into account in $\Delta \hat n_i$.  
Fits are obtained for both $s=\pm1$. The total systematic uncertainty on $b_z$ and $K_3$ is then estimated by adding the deviations of the $s=\pm1$ fits from the original fit in Eq.~\eqref{eq:chi2} in quadrature. 

\begin{figure}
    \centering
    \includegraphics[scale=0.45]{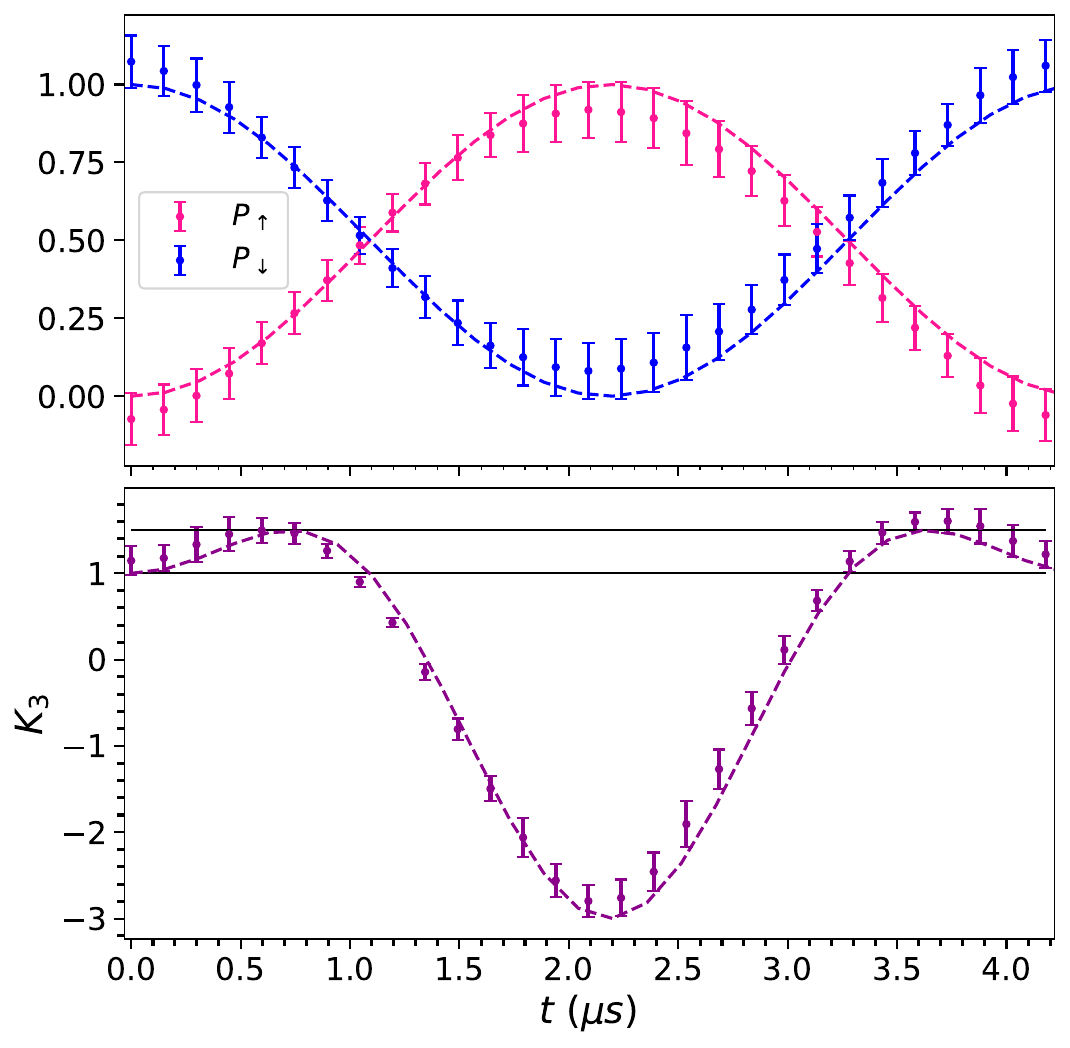}
    \caption{Upper panel: The conditional probability (\ref{eq:Ppm}) of finding a spin up/down muon when the muon is spin down at $t=0$. Lower panel: The Leggett-Garg variable $K_3$~\eqref{eq:K3Deltat} constructed from the data in the upper panel. Theoretical curves for the probabilities and $K_3$ are given by the dashed lines. 
    }
    \label{fig:fittedPandK3}
\end{figure}

When choosing a starting time $t_0$ at which muons are spin-down, the conditional probabilities of finding spin-up and spin-down muons at time $t$ are shown in the upper panel of Fig.~\ref{fig:fittedPandK3}. 
With the conditional probability, we calculate the Leggett-Garg variable $K_3$ from Eqs.~\eqref{eq:K3Deltat} and~\eqref{eq:CfromConditionalP} as shown in the lower panel of Fig.~\ref{fig:fittedPandK3}.
As the violation of LGI arises from the quantum mechanical muon spin precession, the time dependence of the variable $K_3$ is similar to that found in other oscillating systems~\cite{Zhou:2015bbe,Fu:2017hky}.
The associated uncertainties shown in Fig.~\ref{fig:fittedPandK3} are dominated by the theory/systematic uncertainties, obtained from the offset method as described above. On the other hand, the statistical uncertainty from the fit in Eq.~\eqref{eq:chi2} is much smaller, with $\Delta K_3^{\rm stat}\simeq 10^{-3}$.
Even after including our conservative estimate of the theory uncertainty, the largest single bin violation of the LGI corresponds to a significance of  $5.5\sigma$. 
Several neighboring time bins also exceed the LGI bound in the region where the quantum mechanical prediction gives $K_3>1$. 
Combining these bins substantially increases the significance of the observed LGI violation, well beyond the largest single bin significance quoted above.

\section{Conclusion and discussion}

The variety of particles, interactions, and experimental settings available in high energy physics experiments makes them rich laboratories for quantum information science. In this work, we have considered the muon $g-2$ experiments, which provide some of the most precise measurements in high energy physics and, in particular, in muon systems. 
The highly polarized muon ensemble, well characterized storage ring environment, and enormous event samples of order $10^{11}$ muon decays available at muon $g-2$ experiments make them promising platforms for precision studies of temporal quantum correlations.

We formulated the steps to reconstruct the LGI in muon spin precession. 
Using existing time-dependent positron energy spectrum data from the Fermilab Muon $g-2$ experiment, we extracted the autocorrelation function of the longitudinal muon spin and used it to determine the Leggett-Garg variable $K_3$.

As a first illustration, we employed a simplified detector response model to describe the measured decay positron spectrum.
The residual differences between the experimental data and our modeling prediction are used to estimate the systematic uncertainties associated with this simplified detector modeling.
Even with this conservative treatment, we find the LGI is violated with a single-bin significance of $5.5\sigma$ using existing data. 
This represents the first observation of the LGI violation with oscillating muons, and also highlights the potential of muon $g-2$ experiments for performing precision quantum information studies.

As a final remark, the dominant uncertainty in our analysis arises from the simplified modeling of detector effects. The reconstruction procedure developed here can be implemented by the experimental collaboration using full detector information and realistic simulations. This would substantially reduce the systematic uncertainties and could bring the total uncertainty on $K_3$ close to the statistical level, $\mathcal{O}(10^{-3})$, as in the precision measurement of the anomalous precession frequency~\cite{Muong-2:2025xyk}. Combining all available oscillation periods would further improve the sensitivity, potentially enabling one of the most precise measurements of LGI violation.

\begin{acknowledgments}
The work of BB and KC is supported by the U.S. Department of Energy under Grant No. DE–SC0007914.
The work of MC is supported by the National Science Foundation Graduate Research Fellowship Program.
\end{acknowledgments}

\bibliography{ref.bib}

@article{Leggett:1985zz,
    author = "Leggett, A. J. and Garg, Anupam",
    title = "{Quantum mechanics versus macroscopic realism: Is the flux there when nobody looks?}",
    doi = "10.1103/PhysRevLett.54.857",
    journal = "Phys. Rev. Lett.",
    volume = "54",
    pages = "857--860",
    year = "1985"
}

@article{Emary:2013wfl,
    author = "Emary, Clive and Lambert, Neill and Nori, Franco",
    title = "{Leggett{\textendash}Garg inequalities}",
    eprint = "1304.5133",
    archivePrefix = "arXiv",
    primaryClass = "quant-ph",
    doi = "10.1088/0034-4885/77/1/016001",
    journal = "Rept. Prog. Phys.",
    volume = "77",
    number = "1",
    pages = "016001",
    year = "2013"
}

@article{PhysRevA.52.R2497,
  title = {Proposed test for realist theories using Rydberg atoms coupled to a high-$Q$ resonator},
  author = {Huelga, S. F. and Marshall, T. W. and Santos, E.},
  journal = {Phys. Rev. A},
  volume = {52},
  issue = {4},
  pages = {R2497--R2500},
  numpages = {0},
  year = {1995},
  month = {Oct},
  publisher = {American Physical Society},
  doi = {10.1103/PhysRevA.52.R2497},
  url = {https://link.aps.org/doi/10.1103/PhysRevA.52.R2497}
}

@article{Waldherr:2011kgt,
    author = "Waldherr, G. and Neumann, P. and Huelga, S. F. and Jelezko, F. and Wrachtrup, J.",
    title = "{Violation of a Temporal Bell Inequality for Single Spins in a Diamond Defect Center}",
    eprint = "1103.4949",
    archivePrefix = "arXiv",
    primaryClass = "quant-ph",
    doi = "10.1103/PhysRevLett.107.090401",
    journal = "Phys. Rev. Lett.",
    volume = "107",
    number = "9",
    pages = "090401",
    year = "2011"
}

@article{Zhou:2015bbe,
    author = "Zhou, Zong-Quan and Huelga, Susana F. and Li, Chuan-Feng and Guo, Guang-Can",
    title = "{Experimental detection of quantum coherent evolution through the violation of Leggett-Garg-type inequalities}",
    eprint = "1209.2176",
    archivePrefix = "arXiv",
    primaryClass = "quant-ph",
    doi = "10.1103/PhysRevLett.115.113002",
    journal = "Phys. Rev. Lett.",
    volume = "115",
    pages = "113002",
    year = "2015"
}

@article{Gangopadhyay:2013aha,
    author = "Gangopadhyay, D. and Home, D. and Roy, A. Sinha",
    title = "{Probing the Leggett-Garg Inequality for Oscillating Neutral Kaons and Neutrinos}",
    eprint = "1304.2761",
    archivePrefix = "arXiv",
    primaryClass = "quant-ph",
    doi = "10.1103/PhysRevA.88.022115",
    journal = "Phys. Rev. A",
    volume = "88",
    number = "2",
    pages = "022115",
    year = "2013"
}

@article{Alok:2014gya,
    author = "Alok, Ashutosh Kumar and Banerjee, Subhashish and Sankar, S. Uma",
    title = "{Quantum correlations in terms of neutrino oscillation probabilities}",
    eprint = "1411.5536",
    archivePrefix = "arXiv",
    primaryClass = "hep-ph",
    doi = "10.1016/j.nuclphysb.2016.05.001",
    journal = "Nucl. Phys. B",
    volume = "909",
    pages = "65--72",
    year = "2016"
}

@article{Banerjee:2015mha,
    author = "Banerjee, Subhashish and Alok, Ashutosh Kumar and Srikanth, R. and Hiesmayr, Beatrix C.",
    title = "{A quantum information theoretic analysis of three flavor neutrino oscillations}",
    eprint = "1508.03480",
    archivePrefix = "arXiv",
    primaryClass = "hep-ph",
    doi = "10.1140/epjc/s10052-015-3717-x",
    journal = "Eur. Phys. J. C",
    volume = "75",
    number = "10",
    pages = "487",
    year = "2015"
}

@article{Formaggio:2016cuh,
    author = "Formaggio, J. A. and Kaiser, D. I. and Murskyj, M. M. and Weiss, T. E.",
    title = "{Violation of the Leggett-Garg Inequality in Neutrino Oscillations}",
    eprint = "1602.00041",
    archivePrefix = "arXiv",
    primaryClass = "quant-ph",
    doi = "10.1103/PhysRevLett.117.050402",
    journal = "Phys. Rev. Lett.",
    volume = "117",
    number = "5",
    pages = "050402",
    year = "2016"
}

@article{Fu:2017hky,
    author = "Fu, Qiang and Chen, Xurong",
    title = "{Testing violation of the Leggett{\textendash}Garg-type inequality in neutrino oscillations of the Daya Bay experiment}",
    eprint = "1705.08601",
    archivePrefix = "arXiv",
    primaryClass = "hep-ph",
    doi = "10.1140/epjc/s10052-017-5371-y",
    journal = "Eur. Phys. J. C",
    volume = "77",
    number = "11",
    pages = "775",
    year = "2017"
}

@article{Gangopadhyay:2017nsn,
    author = "Gangopadhyay, Debashis and Roy, Animesh Sinha",
    title = "{Three-flavoured neutrino oscillations and the Leggett{\textendash}Garg inequality}",
    eprint = "1702.04646",
    archivePrefix = "arXiv",
    primaryClass = "quant-ph",
    doi = "10.1140/epjc/s10052-017-4837-2",
    journal = "Eur. Phys. J. C",
    volume = "77",
    number = "4",
    pages = "260",
    year = "2017"
}

@article{Naikoo:2019eec,
    author = "Naikoo, Javid and Kumar Alok, Ashutosh and Banerjee, Subhashish and Uma Sankar, S.",
    title = "{Leggett-Garg inequality in the context of three flavour neutrino oscillation}",
    eprint = "1901.10859",
    archivePrefix = "arXiv",
    primaryClass = "hep-ph",
    doi = "10.1103/PhysRevD.99.095001",
    journal = "Phys. Rev. D",
    volume = "99",
    number = "9",
    pages = "095001",
    year = "2019"
}

@article{Wang:2022tnr,
    author = "Wang, Xing-Zhi and Ma, Bo-Qiang",
    title = "{New test of neutrino oscillation coherence with Leggett{\textendash}Garg inequality}",
    eprint = "2201.10597",
    archivePrefix = "arXiv",
    primaryClass = "quant-ph",
    doi = "10.1140/epjc/s10052-022-10053-1",
    journal = "Eur. Phys. J. C",
    volume = "82",
    number = "2",
    pages = "133",
    year = "2022"
}

@article{Alam:2026bxn,
    author = "Alam, Murshed and Brdar, Vedran and Chattopadhyay, Dibya S.",
    title = "{Exploring Quantumness at Long-Baseline Neutrino Experiments}",
    eprint = "2601.15375",
    archivePrefix = "arXiv",
    primaryClass = "hep-ph",
    month = "1",
    year = "2026"
}

@article{Muong-2:2015xgu,
    author = "Grange, J. and others",
    collaboration = "Muon g-2",
    title = "{Muon (g-2) Technical Design Report}",
    eprint = "1501.06858",
    archivePrefix = "arXiv",
    primaryClass = "physics.ins-det",
    reportNumber = "FERMILAB-FN-0992-E, FERMILAB-DESIGN-2014-02",
    doi = "10.2172/1251172",
    month = "1",
    year = "2015"
}

@article{Muong-2:2025xyk,
    author = "Aguillard, D. P. and others",
    collaboration = "Muon g-2",
    title = "{Measurement of the Positive Muon Anomalous Magnetic Moment to 127~ppb}",
    eprint = "2506.03069",
    archivePrefix = "arXiv",
    primaryClass = "hep-ex",
    reportNumber = "FERMILAB-PUB-25-0364-PPD",
    doi = "10.1103/7clf-sm2v",
    journal = "Phys. Rev. Lett.",
    volume = "135",
    number = "10",
    pages = "101802",
    year = "2025"
}

@phdthesis{LaBounty:2024epq,
    author = "LaBounty, Joshua",
    title = "{Analysis of the Anomalous Spin Precession of the Muon for the Fermilab Muon g-2 Experiment}",
    url = "http://hdl.handle.net/1773/51398",
    year = "2024"
}

@article{PhysicsPhysiqueFizika.1.195,
    author = "Bell, J. S.",
    title = "{On the Einstein-Podolsky-Rosen paradox}",
    reportNumber = "RX-1376",
    doi = "10.1103/PhysicsPhysiqueFizika.1.195",
    journal = "Physics Physique Fizika",
    volume = "1",
    pages = "195--200",
    year = "1964"
}

@article{Clauser:1969ny,
    author = "Clauser, John F. and Horne, Michael A. and Shimony, Abner and Holt, Richard A.",
    title = "{Proposed experiment to test local hidden variable theories}",
    doi = "10.1103/PhysRevLett.23.880",
    journal = "Phys. Rev. Lett.",
    volume = "23",
    pages = "880--884",
    year = "1969"
}

@article{HORODECKI1997333,
    author = "Horodecki, Pawel",
    title = "{Separability criterion and inseparable mixed states with positive partial transposition}",
    eprint = "quant-ph/9703004",
    archivePrefix = "arXiv",
    doi = "10.1016/S0375-9601(97)00416-7",
    journal = "Phys. Lett. A",
    volume = "232",
    pages = "333",
    year = "1997"
}

@article{PhysRevLett.80.2245,
    author = "Wootters, William K.",
    title = "{Entanglement of formation of an arbitrary state of two qubits}",
    eprint = "quant-ph/9709029",
    archivePrefix = "arXiv",
    doi = "10.1103/PhysRevLett.80.2245",
    journal = "Phys. Rev. Lett.",
    volume = "80",
    pages = "2245--2248",
    year = "1998"
}

@article{Afik:2020onf,
    author = "Afik, Yoav and de Nova, Juan Ram\'on Mu\~noz",
    title = "{Entanglement and quantum tomography with top quarks at the LHC}",
    eprint = "2003.02280",
    archivePrefix = "arXiv",
    primaryClass = "quant-ph",
    doi = "10.1140/epjp/s13360-021-01902-1",
    journal = "Eur. Phys. J. Plus",
    volume = "136",
    number = "9",
    pages = "907",
    year = "2021"
}

@article{Fabbrichesi:2021npl,
    author = "Fabbrichesi, M. and Floreanini, R. and Panizzo, G.",
    title = "{Testing Bell Inequalities at the LHC with Top-Quark Pairs}",
    eprint = "2102.11883",
    archivePrefix = "arXiv",
    primaryClass = "hep-ph",
    doi = "10.1103/PhysRevLett.127.161801",
    journal = "Phys. Rev. Lett.",
    volume = "127",
    number = "16",
    pages = "161801",
    year = "2021"
}

@article{Severi:2021cnj,
    author = "Severi, Claudio and Boschi, Cristian Degli Esposti and Maltoni, Fabio and Sioli, Maximiliano",
    title = "{Quantum tops at the LHC: from entanglement to Bell inequalities}",
    eprint = "2110.10112",
    archivePrefix = "arXiv",
    primaryClass = "hep-ph",
    doi = "10.1140/epjc/s10052-022-10245-9",
    journal = "Eur. Phys. J. C",
    volume = "82",
    number = "4",
    pages = "285",
    year = "2022"
}

@article{Afik:2022kwm,
    author = "Afik, Yoav and de Nova, Juan Ram\'on Mu\~noz",
    title = "{Quantum information with top quarks in QCD}",
    eprint = "2203.05582",
    archivePrefix = "arXiv",
    primaryClass = "quant-ph",
    doi = "10.22331/q-2022-09-29-820",
    journal = "Quantum",
    volume = "6",
    pages = "820",
    year = "2022"
}

@article{Afik:2022dgh,
    author = "Afik, Yoav and de Nova, Juan Ram\'on Mu\~noz",
    title = "{Quantum Discord and Steering in Top Quarks at the LHC}",
    eprint = "2209.03969",
    archivePrefix = "arXiv",
    primaryClass = "quant-ph",
    doi = "10.1103/PhysRevLett.130.221801",
    journal = "Phys. Rev. Lett.",
    volume = "130",
    number = "22",
    pages = "221801",
    year = "2023"
}

@article{Aguilar_Saavedra_2022,
    author = "Aguilar-Saavedra, J. A. and Casas, J. A.",
    title = "{Improved tests of entanglement and Bell inequalities with LHC tops}",
    eprint = "2205.00542",
    archivePrefix = "arXiv",
    primaryClass = "hep-ph",
    reportNumber = "IFT-UAM/CSIC-22-45",
    doi = "10.1140/epjc/s10052-022-10630-4",
    journal = "Eur. Phys. J. C",
    volume = "82",
    number = "8",
    pages = "666",
    year = "2022"
}

@article{Dong_2024,
    author = "Dong, Zhongtian and Gon\c{c}alves, Dorival and Kong, Kyoungchul and Navarro, Alberto",
    title = "{Entanglement and Bell inequalities with boosted tt\textasciimacron{}}",
    eprint = "2305.07075",
    archivePrefix = "arXiv",
    primaryClass = "hep-ph",
    doi = "10.1103/PhysRevD.109.115023",
    journal = "Phys. Rev. D",
    volume = "109",
    number = "11",
    pages = "115023",
    year = "2024"
}

@article{Han:2023fci,
    author = "Han, Tao and Low, Matthew and Wu, Tong Arthur",
    title = "{Quantum entanglement and Bell inequality violation in semi-leptonic top decays}",
    eprint = "2310.17696",
    archivePrefix = "arXiv",
    primaryClass = "hep-ph",
    reportNumber = "PITT-PACC-2316",
    doi = "10.1007/JHEP07(2024)192",
    journal = "JHEP",
    volume = "07",
    pages = "192",
    year = "2024"
}

@article{Cheng:2023qmz,
    author = "Cheng, Kun and Han, Tao and Low, Matthew",
    title = "{Optimizing fictitious states for Bell inequality violation in bipartite qubit systems with applications to the tt{\textasciimacron} system}",
    eprint = "2311.09166",
    archivePrefix = "arXiv",
    primaryClass = "hep-ph",
    reportNumber = "PITT-PACC-2321",
    doi = "10.1103/PhysRevD.109.116005",
    journal = "Phys. Rev. D",
    volume = "109",
    number = "11",
    pages = "116005",
    year = "2024"
}

@article{Cheng:2024btk,
    author = "Cheng, Kun and Han, Tao and Low, Matthew",
    title = "{Optimizing entanglement and Bell inequality violation in top antitop events}",
    eprint = "2407.01672",
    archivePrefix = "arXiv",
    primaryClass = "hep-ph",
    reportNumber = "PITT-PACC-2401",
    doi = "10.1103/PhysRevD.111.033004",
    journal = "Phys. Rev. D",
    volume = "111",
    number = "3",
    pages = "033004",
    year = "2025"
}

@article{Cheng:2024rxi,
    author = "Cheng, Kun and Han, Tao and Low, Matthew",
    title = "{Quantum tomography at colliders: With or without decays}",
    eprint = "2410.08303",
    archivePrefix = "arXiv",
    primaryClass = "hep-ph",
    reportNumber = "PITT-PACC-2408",
    doi = "10.1016/j.physletb.2025.139675",
    journal = "Phys. Lett. B",
    volume = "868",
    pages = "139675",
    year = "2025"
}

@article{Cheng:2025zcf,
    author = "Cheng, Kun and Han, Tao and Low, Matthew and Wu, Tong Arthur",
    title = "{Quantum Tomography in Neutral Meson and Antimeson Systems}",
    eprint = "2507.12513",
    archivePrefix = "arXiv",
    primaryClass = "hep-ph",
    doi = "10.1103/t9vg-hk6c",
    journal = "Phys. Rev. Lett.",
    volume = "136",
    number = "18",
    pages = "181803",
    year = "2026"
}

@article{CMSCollaboration_2024,
    author = "Hayrapetyan, Aram and others",
    collaboration = "CMS",
    title = "{Observation of quantum entanglement in top quark pair production in proton\textendash{}proton collisions at $\sqrt{s} = 13$ TeV}",
    eprint = "2406.03976",
    archivePrefix = "arXiv",
    primaryClass = "hep-ex",
    reportNumber = "CMS-TOP-23-001, CERN-EP-2024-137",
    doi = "10.1088/1361-6633/ad7e4d",
    journal = "Rept. Prog. Phys.",
    volume = "87",
    number = "11",
    pages = "117801",
    year = "2024"
}

@article{ATLAS:2023fsd,
    author = "Aad, Georges and others",
    collaboration = "ATLAS",
    title = "{Observation of quantum entanglement with top quarks at the ATLAS detector}",
    eprint = "2311.07288",
    archivePrefix = "arXiv",
    primaryClass = "hep-ex",
    reportNumber = "CERN-EP-2023-230",
    doi = "10.1038/s41586-024-07824-z",
    journal = "Nature",
    volume = "633",
    number = "8030",
    pages = "542--547",
    year = "2024"
}

@article{Han:2024ugl,
    author = "Han, Tao and Low, Matthew and McGinnis, Navin and Su, Shufang",
    title = "{Measuring quantum discord at the LHC}",
    eprint = "2412.21158",
    archivePrefix = "arXiv",
    primaryClass = "hep-ph",
    reportNumber = "PITT-PACC-2316",
    doi = "10.1007/JHEP05(2025)081",
    journal = "JHEP",
    volume = "05",
    pages = "081",
    year = "2025"
}

@article{Altakach:2022ywa,
    author = "Altakach, Mohammad Mahdi and Lamba, Priyanka and Maltoni, Fabio and Mawatari, Kentarou and Sakurai, Kazuki",
    title = "{Quantum information and CP measurement in H{\textrightarrow}{\ensuremath{\tau}}+{\ensuremath{\tau}}- at future lepton colliders}",
    eprint = "2211.10513",
    archivePrefix = "arXiv",
    primaryClass = "hep-ph",
    doi = "10.1103/PhysRevD.107.093002",
    journal = "Phys. Rev. D",
    volume = "107",
    number = "9",
    pages = "093002",
    year = "2023"
}

@article{Ehataht:2023zzt,
    author = {Ehat{\"a}ht, Karl and Fabbrichesi, Marco and Marzola, Luca and Veelken, Christian},
    title = "{Probing entanglement and testing Bell inequality violation with e+e-{\textrightarrow}{\ensuremath{\tau}}+{\ensuremath{\tau}}- at Belle II}",
    eprint = "2311.17555",
    archivePrefix = "arXiv",
    primaryClass = "hep-ph",
    doi = "10.1103/PhysRevD.109.032005",
    journal = "Phys. Rev. D",
    volume = "109",
    number = "3",
    pages = "032005",
    year = "2024"
}

@article{Ma:2023yvd,
    author = "Ma, Kai and Li, Tong",
    title = "{Testing Bell inequality through ${\boldsymbol h{\bf\rightarrow}\boldsymbol\tau\boldsymbol\tau }$ at CEPC*}",
    eprint = "2309.08103",
    archivePrefix = "arXiv",
    primaryClass = "hep-ph",
    doi = "10.1088/1674-1137/ad62d8",
    journal = "Chin. Phys. C",
    volume = "48",
    number = "10",
    pages = "103105",
    year = "2024"
}

@article{Fabbrichesi:2024wcd,
    author = "Fabbrichesi, M. and Marzola, L.",
    title = "{Quantum tomography with {\ensuremath{\tau}} leptons at the FCC-ee: Entanglement, Bell inequality violation, sin{\ensuremath{\theta}}W, and anomalous couplings}",
    eprint = "2405.09201",
    archivePrefix = "arXiv",
    primaryClass = "hep-ph",
    doi = "10.1103/PhysRevD.110.076004",
    journal = "Phys. Rev. D",
    volume = "110",
    number = "7",
    pages = "076004",
    year = "2024"
}

@article{Han:2025ewp,
    author = "Han, Tao and Low, Matthew and Su, Youle",
    title = "{Entanglement and Bell nonlocality in {\ensuremath{\tau}}$^{+}${\ensuremath{\tau}}$^{−}$ at the BEPC}",
    eprint = "2501.04801",
    archivePrefix = "arXiv",
    primaryClass = "hep-ph",
    reportNumber = "PITT-PACC-2412",
    doi = "10.1007/JHEP10(2025)217",
    journal = "JHEP",
    volume = "10",
    pages = "217",
    year = "2025"
}

@article{Zhang:2025mmm,
    author = "Zhang, Yulei and Zhou, Bai-Hong and Liu, Qi-Bin and Li, Shu and Hsu, Shih-Chieh and Han, Tao and Low, Matthew and Wu, Tong Arthur",
    title = "{Entanglement and Bell Nonlocality in $\tau^+ \tau^-$ at the LHC using Machine Learning for Neutrino Reconstruction}",
    eprint = "2504.01496",
    archivePrefix = "arXiv",
    primaryClass = "hep-ph",
    month = "4",
    year = "2025"
}

@article{Barr:2021zcp,
    author = "Barr, Alan J.",
    title = "{Testing Bell inequalities in Higgs boson decays}",
    eprint = "2106.01377",
    archivePrefix = "arXiv",
    primaryClass = "hep-ph",
    doi = "10.1016/j.physletb.2021.136866",
    journal = "Phys. Lett. B",
    volume = "825",
    pages = "136866",
    year = "2022"
}

@article{Barr:2022wyq,
    author = "Barr, Alan J. and Caban, Pawel and Rembieli\'nski, Jakub",
    title = "{Bell-type inequalities for systems of relativistic vector bosons}",
    eprint = "2204.11063",
    archivePrefix = "arXiv",
    primaryClass = "quant-ph",
    doi = "10.22331/q-2023-07-27-1070",
    journal = "Quantum",
    volume = "7",
    pages = "1070",
    year = "2023"
}

@article{Ashby-Pickering:2022umy,
    author = "Ashby-Pickering, Rachel and Barr, Alan J. and Wierzchucka, Agnieszka",
    title = "{Quantum state tomography, entanglement detection and Bell violation prospects in weak decays of massive particles}",
    eprint = "2209.13990",
    archivePrefix = "arXiv",
    primaryClass = "quant-ph",
    doi = "10.1007/JHEP05(2023)020",
    journal = "JHEP",
    volume = "05",
    pages = "020",
    year = "2023"
}

@article{Aguilar-Saavedra:2022wam,
    author = "Aguilar-Saavedra, J. A. and Bernal, A. and Casas, J. A. and Moreno, J. M.",
    title = "{Testing entanglement and Bell inequalities in H\textrightarrow{}ZZ}",
    eprint = "2209.13441",
    archivePrefix = "arXiv",
    primaryClass = "hep-ph",
    doi = "10.1103/PhysRevD.107.016012",
    journal = "Phys. Rev. D",
    volume = "107",
    number = "1",
    pages = "016012",
    year = "2023"
}

@article{Fabbrichesi:2023cev,
    author = "Fabbrichesi, Marco and Floreanini, Roberto and Gabrielli, Emidio and Marzola, Luca",
    title = "{Bell inequalities and quantum entanglement in weak gauge boson production at the LHC and future colliders}",
    eprint = "2302.00683",
    archivePrefix = "arXiv",
    primaryClass = "hep-ph",
    doi = "10.1140/epjc/s10052-023-11935-8",
    journal = "Eur. Phys. J. C",
    volume = "83",
    number = "9",
    pages = "823",
    year = "2023"
}

@article{Fabbri:2023ncz,
    author = "Fabbri, Federica and Howarth, James and Maurin, Theo",
    title = "{Isolating semi-leptonic $H\rightarrow WW^{*}$decays for Bell inequality tests}",
    eprint = "2307.13783",
    archivePrefix = "arXiv",
    primaryClass = "hep-ph",
    doi = "10.1140/epjc/s10052-023-12371-4",
    journal = "Eur. Phys. J. C",
    volume = "84",
    number = "1",
    pages = "20",
    year = "2024"
}

@article{Bi:2023uop,
    author = "Bi, Qi and Cao, Qing-Hong and Cheng, Kun and Zhang, Hao",
    title = "{New observables for testing Bell inequalities in W boson pair production}",
    eprint = "2307.14895",
    archivePrefix = "arXiv",
    primaryClass = "hep-ph",
    doi = "10.1103/PhysRevD.109.036022",
    journal = "Phys. Rev. D",
    volume = "109",
    number = "3",
    pages = "036022",
    year = "2024"
}

@article{Morales:2023gow,
    author = "Morales, R. A.",
    title = "{Exploring Bell inequalities and quantum entanglement in vector boson scattering}",
    eprint = "2306.17247",
    archivePrefix = "arXiv",
    primaryClass = "hep-ph",
    doi = "10.1140/epjp/s13360-023-04784-7",
    journal = "Eur. Phys. J. Plus",
    volume = "138",
    number = "12",
    pages = "1157",
    year = "2023"
}

@article{Bernal:2024xhm,
    author = "Bernal, Alexander and Caban, Pawe{\l} and Rembieli{\'n}ski, Jakub",
    title = "{Entanglement and Bell inequality violation in vector diboson systems produced in decays of spin-0 particles}",
    eprint = "2405.16525",
    archivePrefix = "arXiv",
    primaryClass = "hep-ph",
    doi = "10.1038/s41598-025-07747-3",
    journal = "Sci. Rep.",
    volume = "15",
    number = "1",
    pages = "23410",
    year = "2025"
}

@article{Grossi:2024jae,
    author = "Grossi, Michele and Pelliccioli, Giovanni and Vicini, Alessandro",
    title = "{From angular coefficients to quantum observables: a phenomenological appraisal in di-boson systems}",
    eprint = "2409.16731",
    archivePrefix = "arXiv",
    primaryClass = "hep-ph",
    reportNumber = "COMETA-2024-24, MPP-2024-183, TIF-UNIMI-2024-15",
    doi = "10.1007/JHEP12(2024)120",
    journal = "JHEP",
    volume = "12",
    pages = "120",
    year = "2024"
}

@article{Goncalves:2025qem,
    author = "Gon{\c{c}}alves, Dorival and Kaladharan, Ajay and Krauss, Frank and Navarro, Alberto",
    title = "{Quantum Entanglement is Quantum: ZZ Production at the LHC}",
    eprint = "2505.12125",
    archivePrefix = "arXiv",
    primaryClass = "hep-ph",
    month = "5",
    year = "2025"
}

@article{Goncalves:2025xer,
    author = "Gon{\c{c}}alves, Dorival and Kaladharan, Ajay and Navarro, Alberto",
    title = "{Higher-Order Corrections to Quantum Observables in $h\to WW^*$}",
    eprint = "2506.19951",
    archivePrefix = "arXiv",
    primaryClass = "hep-ph",
    month = "6",
    year = "2025"
}

@article{Cheng:2025zaw,
    author = "Cheng, Kun and Han, Tao and Trifinopoulos, Sokratis",
    title = "{Quantum Information at the Electron-Ion Collider}",
    eprint = "2510.23773",
    archivePrefix = "arXiv",
    primaryClass = "hep-ph",
    reportNumber = "PITT-PACC-2509, CERN-TH-2025-205, MIT-CTP/5943",
    month = "10",
    year = "2025"
}

@article{Ai:2025wnt,
    author = "Ai, Tengyu and Bi, Qi and He, Yuxin and Liu, Jia and Wang, Xiao-Ping",
    title = "{Ultimate Quantum Precision Limit at Colliders: Conditions and Case Studies}",
    eprint = "2506.10673",
    archivePrefix = "arXiv",
    primaryClass = "hep-ph",
    reportNumber = "CPTNP-2025-019",
    doi = "10.1103/3m4t-pk9b",
    journal = "Phys. Rev. Lett.",
    volume = "135",
    number = "24",
    pages = "241804",
    year = "2025"
}

@article{Afik:2025grr,
    author = "Afik, Yoav and Kats, Yevgeny and de Nova, Juan Ram{\'o}n Mu{\~n}oz and Soffer, Abner and Uzan, David",
    title = "{Entanglement and Bell nonlocality with bottom-quark pairs at hadron colliders}",
    eprint = "2406.04402",
    archivePrefix = "arXiv",
    primaryClass = "hep-ph",
    doi = "10.1103/fhkc-kfhr",
    journal = "Phys. Rev. D",
    volume = "111",
    number = "11",
    pages = "L111902",
    year = "2025"
}

@article{Fang:2026ddi,
    author = "Fang, Yi-Jing and Bhoonah, Amit and Cheng, Kun and Han, Tao and Liu, Yandong and Zhang, Hao",
    title = "{Spin Correlation and Quantum Entanglement of Fermion Pairs in Transversely Polarized $e^-e^+$ Collisions}",
    eprint = "2604.11887",
    archivePrefix = "arXiv",
    primaryClass = "hep-ph",
    reportNumber = "PITT-PACC-2604",
    month = "4",
    year = "2026"
}

@article{Guo:2026yhz,
    author = "Guo, Yu-Chen and Han, Tao and Low, Matthew and Su, Youle",
    title = "{Quantum Tomography of Fermion Pairs in $e^+e^-$ Collisions: Longitudinal Beam Polarization Effects}",
    eprint = "2602.02719",
    archivePrefix = "arXiv",
    primaryClass = "hep-ph",
    month = "2",
    year = "2026"
}

@article{Lin:2025eci,
    author = "Lin, Shi-Jia and Liu, Ming-Jun and Shao, Ding Yu and Wei, Shu-Yi",
    title = "{Spin correlations and Bell nonlocality in $ \Lambda \overline{\Lambda} $ pair production from e$^{+}$e$^{−}$ collisions with a thrust cut}",
    eprint = "2507.15387",
    archivePrefix = "arXiv",
    primaryClass = "hep-ph",
    doi = "10.1007/JHEP11(2025)082",
    journal = "JHEP",
    volume = "11",
    pages = "082",
    year = "2025"
}
\bibliographystyle{apsrev4-1}

\end{document}